\documentclass{aa}
\usepackage{txfonts}
\usepackage{epsfig}
\usepackage{natbib}
\bibpunct{(}{)}{;}{a}{}{,} 

\def\0{\hspace*{0.5em}}
\begin{document}


\title{Discovery of a magnetic field in the early B-type star {\boldmath $\sigma$} Lupi\thanks{Based on observations obtained
by the International Ultraviolet Explorer, collected at NASA Goddard Space Flight Center and Villafranca Satellite Tracking Station of the European Space Agency.
Optical observations have been obtained at the Anglo Australian Telescope (AAT) and at the Canada-France-Hawaii              
Telescope (CFHT) which is operated by the National Research             
Council of Canada, the Institut National des Sciences de l'Univers      
of the Centre National de la Recherche Scientifique of France,            
and the University of Hawaii. }}
\author{H.F. Henrichs\inst{1,2}  \and K. Kolenberg\inst{3,4} \and B. Plaggenborg\inst{1}
\and S.C. Marsden\inst{5,6} \and I.A. Waite\inst{7} \and J.D. Landstreet\inst{8,9} \and G.A. Wade\inst{10} \and  J.H.~Grunhut\inst{10}  \and M.E.~Oksala\inst{11}  \and the MiMeS collaboration
}

\institute{
Astronomical Institute 'Anton Pannekoek', University of Amsterdam,
Science Park 904, 1098 XH Amsterdam, Netherlands
\and Department of Astrophysics/IMAPP
Radboud University Nijmegen, P.O. Box 9010, 6500 GL Nijmegen, Netherlands
\and Harvard-Smithsonian Center for Astrophysics, 60 Garden Street,
Cambridge MA 02138, USA
\and Instituut voor Sterrenkunde, K. U. Leuven, Celestijnenlaan 200D, 3001
Leuven, Belgium
\and Australian Astronomical Observatory, PO Box 296, Epping, NSW 1710, Australia
\and Centre for Astronomy, School of Engineering and Physical Sciences, James Cook University, Townsville, 4811, Australia
\and Faculty of Sciences, University of Southern Queensland, Toowoomba, Qld 4350 Australia
\and Dept.\ of Physics and Astronomy, University of Western Ontario, London, ON N6A 3K7 Canada 
\and Armagh Observatory, College Hill, Armagh, Northern Ireland BT61 9DG
\and Dept.\ of Physics, Royal Military College of Canada, PO Box 17000, Station Forces, Kingston, Ontario, Canada
\and Dept.\ of Physics and Astronomy, University of Delaware, Newark, DE, USA 
}

\offprints{H.F. Henrichs,
\email{h.f.henrichs@uva.nl}}

\date{Received date / Accepted date}

\keywords{Stars: early-type -- Stars: magnetic fields -- Stars: winds, outflows -- Stars: rotation -- Stars: starspots -- Stars: abundances}

\abstract
{Magnetic early B-type stars are rare. Indirect indicators are needed to identify them before investing in time-intensive spectropolarimetric observations.
}
{We use the strongest indirect indicator of a magnetic field in B stars, which is periodic variability of ultraviolet (UV) stellar wind lines occurring symmetric about the approximate rest wavelength.
Our aim is to identify probable magnetic candidates which would become targets for follow-up spectropolarimetry to search for a magnetic field.
}
{From the UV wind line variability the B1/B2V star $\sigma$ Lupi emerged as a new magnetic candidate star. AAT spectropolarimetric measurements with SEMPOL were obtained. The longitudinal component of the magnetic field integrated over the visible surface of the star was determined with the Least-Squares Deconvolution method.
}
{The UV line variations of $\sigma$ Lupi are similar to what is known in magnetic B stars, but no periodicity could be determined. We detected a varying longitudinal magnetic field with amplitude of about 100 G with error bars of typically 20 G, which supports an oblique magnetic-rotator configuration. The equivalent width variations of the UV lines, the magnetic and the optical-line variations are consistent with the photometric period of 3.02 d, which we identify with the rotation period of the star.  Additional observations with ESPaDOnS attached to the CFHT confirmed this discovery, and allowed the determination of a precise magnetic period. Analysis revealed that $\sigma$ Lupi is a helium-strong star, with an enhanced nitrogen abundance and an underabundance of carbon, and has a chemically spotted surface.
}
{$\sigma$ Lupi is a magnetic oblique rotator, and is a He-strong star. Like in other magnetic B stars the UV wind emission appears to originate close to the magnetic equatorial plane, with maximum emission occurring when a magnetic pole points towards the Earth. The $3.01972\pm 0.00043$~d magnetic rotation period is consistent with the photometric period, with maximum light corresponding to maximum magnetic field.
}

\titlerunning{The magnetic early B-type star $\sigma$~Lupi}

\authorrunning{H.~F. Henrichs, K. Kolenberg, B. Plaggenborg et al.}

\maketitle

\section{Introduction}
The star \object{$\sigma$~Lupi} (\object{HD 127381}, \object{HR 5425}, \object{HIP 71121}) has been classified as spectral type B2III by \cite{hiltner:1969}. Other classifications include B2V \citep{devaucouleurs:1957} and B2III/IV \citep{houk:1978}. Recently, \cite{levenhagen:2006} concluded that B1/B2V would be more appropriate, based on the photospheric parameters evaluated in their work.

Although \cite{shobbrook:1978} found no photometric variations, \cite{vanderlinden:1987} reported a significant amplitude in variations in the V band of 0.01 mag. This emerged during a photometric campaign in which $\sigma$ Lup was used as a comparison star.
In a follow-up study \cite{jerzykiewicz:1992} collected about one-hundred $uvby$ observations in 1986 and found two significant peaks in the periodogram: at $P$ = 3.02 d and at an alias period of 1.49 d. They also analysed the previous dataset of \cite{vanderlinden:1987}, consisting of 22 observations between 1983 and 1984, and found the same periodicity, allowing them to improve the accuracy of the main period. \cite{jerzykiewicz:1992} concluded that the star may be an ellipsoidal variable, but that the 3.02 d period could not be ruled out to be caused by rotation or by high overtone $g$-mode pulsations. They found that rotational modulation is a reasonable working hypothesis for the star. So far there exists no pulsation study or abundance analysis. The star has been found to be a photometric variable in the {\sl Hipparcos} data \citep{koen:2002} with a frequency of 10.93482 d$^{-1}$ and amplitude 0.0031 magnitude in V, based on a sinusoid fit to the 143 datapoints. The nature of this periodicity is not clear. 

\begin{figure*}[!ht]
\epsfig{file=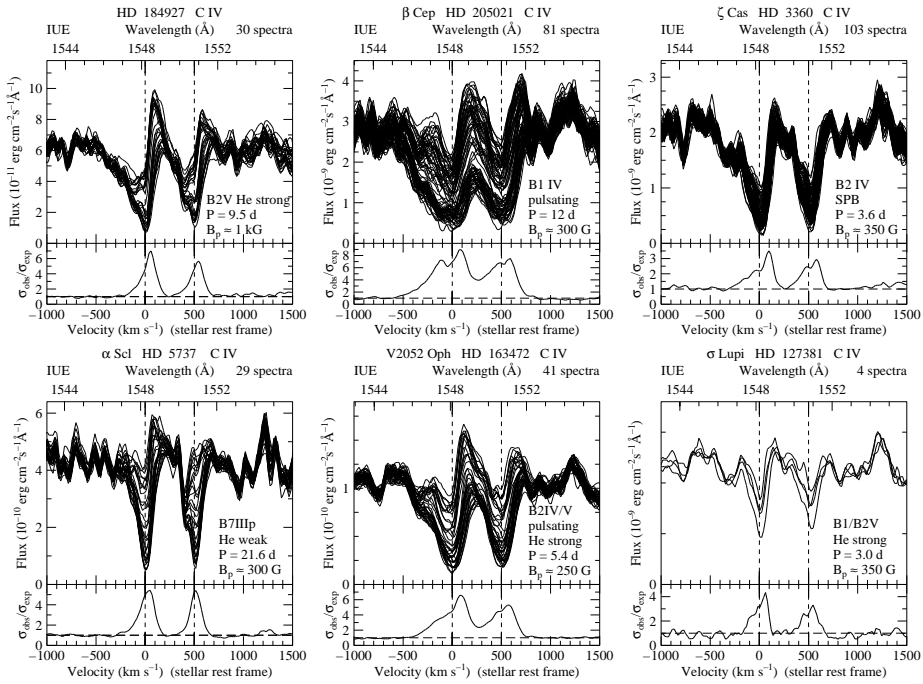,width=1.37\columnwidth,clip=}
\caption{Gallery of magnetic B stars with signatures of magnetically-confined stellar winds
in the \ion{C}{iv} line. The typical modulation in the known magnetic B stars, the He-strong star HD 184927 (top left) and the He-weak star $\alpha$ Scl (bottom left) is very similar to that observed in the four stars $\beta$ Cep, V2052 Oph, $\zeta$ Cas (see text for references) and $\sigma$ Lup (this paper), which led to the discoveries of their magnetic fields. Note that rotation period, pulsation properties and helium peculiarity differ.  In each figure the upper panel shows an overplot of all
available International Ultraviolet Explorer (IUE) spectra, taken over many
rotational cycles. The lower panel displays the ratio of the observed
variation to the expected variation (due to the noise), showing the
velocity range in the stellar rest frame within which significant
variations occur. Note that the whole profile moves up and down,
approximately symmetrically with respect to zero velocity. This phenomenon is only observed in magnetic early-type stars.}
\label{fig:civ}
\end{figure*}

In the course of our investigation of the role of magnetic fields in massive stars a number of indirect magnetic indicators have been identified \citep{henrichs:2005}. The most significant indicator is a strictly periodic variation in the well-known UV stellar wind resonance lines, in particular \ion{C}{iv}, \ion{Si}{iv}, \ion{N}{v} and \ion{Al}{iii}. The variation observed in these lines is, in these magnetic objects, restricted to a velocity interval symmetric around the rest wavelength, in contrast to the often observed high-velocity absorption variations seen in many O and B stars (see e.g.\ \citealt{kaper:1997,fullerton:2003}). This symmetric low-velocity variability in the UV wind lines of B stars is encountered among all studied magnetic oblique rotators (see examples in Fig.~\ref{fig:civ}) which include many helium peculiar stars with spectral type around B1/B2 for the He-strong stars and around B7 for the He-weak stars. Prime examples are the He-strong 
B2V star \object{HD 184927} where the varying UV lines (repeated in Fig.~\ref{fig:civ}) were discovered by \cite{barker:1986},  and the He weak B7IIIp star \object{$\alpha$ Scl}, of which the rotation period was discovered by \cite{shore:1990} from UV data (repeated in Fig.~\ref{fig:civ}). To the group of early B-type stars with this same specific UV variations also belong \object{$\beta$ Cep} (B1IV), \object{V2052 Oph} (B2IV-V) and \object{$\zeta$~Cas} (B2IV): respectively a $\beta$ Cephei variable, a He-strong $\beta$ Cephei star and an SPB (slowly pulsating B) star.  The UV wind lines of these three stars and of the magnetic B stars were recognised to behave very similarly, which prompted a search for a magnetic field in these stars. After the prediction by \cite{henrichs:1993}, all three were indeed found to be magnetic \citep{henrichs:2000a, henrichs:2011, neiner:2003a, neiner:2003c}, based on observations obtained with the Musicos spectropolarimeter \citep{donati:1999}. The dipolar field strengths at the equator of these stars are less than 1000 G. It should be noted that the very accurately determined rotation periods of these stars cover a rather wide range: 12.00 d, 5.37 d and 3.64 d for $\beta$ Cep, V2052 Oph and $\zeta$~Cas, respectively.

\cite{grady:1987} already noted \ion{C}{iv} variations in IUE spectra of $\sigma$~Lup similar to those in $\zeta$~Cas (long before the magnetic origin of this behaviour was recognised), but it appears now that the two spectra on which this conclusion was based were severely overexposed, inhibiting reliable quantitative results. 

No X-ray or radio flux from $\sigma$~Lup has been detected, both indicators of a magnetic environment around the star.
An upper limit of 10$^{30.38}$ erg\,s$^{-1}$ on the X-ray luminosity was obtained with the ROSAT satellite \citep{berghofer:1996}.

In this paper we report the analysis of the stellar wind lines of $\sigma$ Lup and show that the variations observed in 4 subsequently taken IUE spectra are without much doubt due to a corotating magnetosphere. In light of the above, we adopt the photometric period of 3.02 d as the rotation period of the star, following the conclusion by \cite{jerzykiewicz:1992}. We reanalyse all available photometric data in Sect.\ \ref{phase}. In Sect.\ 4 we review the UV results. In Sects.\ \ref{magneticobs} and \ref{magneticresults} we report on magnetic follow-up measurements of this star which confirmed our conjecture. In Sect.\ \ref{ew} we investigate the variations in equivalent width of selected spectral lines. The spectra allowed a subsequent abundance analysis, which we present in Sect.\ \ref{abund}. In the last section we summarise and discuss our conclusions.

\section{Stellar properties}
\label{star}

\begin{table}[b!]
\caption[Stellar parameters]{Adopted and derived stellar parameters for \object{$\sigma$ Lup.}}
\label{param}
\begin{tabular}{lll}
\hline
\hline
& & Reference\\
\hline
Spectral Type                & B1/B2V                & 1 \\
$V$                          & 4.416                 & 2 \\
$d$ (pc)                     & 176$^{+26}_{-20}$     & 3  \\
log($L$/$L_{\odot}$)         & 3.76 $\pm$ 0.06       & 1 \\
$T_{\rm eff}$ (K)            & 23 000 $\pm$ 550      & 1 \\
log\,$g$ (cm\,s$^{-2}$)      & 4.02 $\pm$ 0.10       & 1 \\
$R/R_{\odot}$                & 4.8 $\pm $ 0.5        & Sect.\ \ref{star} \\
$M/M_{\odot}$                & 9.0 $\pm$ 0.5         & 1 \\
log Age (yr)                 & 7.13 $\pm$ 0.13       & 1 \\
$v$sin$i$ (km\,s$^{-1}$)     & 68 $\pm$ 6            & Sect.\ \ref{abund} \\
$P_{\rm phot}$ (d)           & 3.01938 $\pm$ 0.00022 & Sect.\ \ref{phase} \\
$P_{\rm magn}$ (d)           & 3.01972 $\pm$ 0.00043 & Sect.\ \ref{magneticresults} \\
$v_{\rm rad}$ (km\,s$^{-1}$) & 0.0 $\pm$ 0.5         & Sect.\ \ref{abund} \\

\hline
\label{tab:parameters}
\end{tabular}\\
1. \cite{levenhagen:2006} -- 2. SIMBAD --
3. \cite{vanleeuwen:2007} -- 4. \cite{jerzykiewicz:1992}
\end{table}

According to the Bright Star Catalogue \citep{bsc5:1991} $\sigma$~Lup is a member of the local association Pleiades group.
The adopted stellar parameters and other relevant observational aspects are collected in Table~\ref{tab:parameters}.
We have adopted the results of the recent analysis based on high-quality spectra by \cite{levenhagen:2006}, rather than the extrapolated and derived values as presented by \cite{prinja:1989}, who adopted a much lower effective temperature (17500~K) and therefore a much larger radius (8.8 R$_{\odot}$) to be consistent with the luminosity (which has a comparable value in the two papers).   \cite{sokolov:1995} derived  $T_{\rm eff}$ =  25 980 $\pm$ 3250~K from the slope of the Balmer continuuum, consistent with the temperature of \cite{levenhagen:2006}. They derived a projected rotational velocity of $v$sin$i$ = 80 $\pm$ 14 km\,s$^{-1}$, whereas \cite{brown:1997} found $v$sin$i = 69 \pm 10$ km\,s$^{-1}$. Our analysis (Sect.\ \ref{abund}) gives very similarly $68 \pm 6$ km\,s$^{-1}$, which we adopt in this paper. With the stellar parameters in Table~\ref{tab:parameters} we derive a radius of the star between 4.25 and 5.37 R$_{\odot}$ which is consistent with the radius of 4.5 R$_{\odot}$ measured by \cite{pasinetti:2001}. This corresponds to the calculated value of $P$/sin$i$ of 3.0 $\pm$ 0.9 d, compatible with the main photometric period, which we identify therefore with the rotational period of the star.

\section{Photometric analysis and new observations}
\label{phase}

\cite{jerzykiewicz:1992} determined the period $P$ = 3.0186 $\pm$ 0.0004 d from the two photometric datasets obtained as part of the Long-term Photometry of Variables (LTPV) programme at ESO \citep{manfroid:1991}.
If we define maximum light as phase zero and take the weighted average of all the reported epochs of maximum light in all four photometric colors of these two datasets, we find for the photometric ephemeris:
\begin{eqnarray}
\nonumber T({\rm LTPV}) = &{\rm HJD}\ 2445603.65\pm0.11 &\\
&+\ n \times(3.0186 \pm 0.0004)&
\label{eq:phase}
\end{eqnarray}

\noindent with $n$ the number of cycles. Since \cite{koen:2002} did not find this period in the {\sl Hipparcos} data, we reconsidered the dataset. By using a CLEAN analysis \citep{roberts:1987} we verified that the 3 d period is indeed not present. However, there are two obvious outliers at HJD 2448309.52 and 2448818.90 which deviate more 86$\sigma$ and 26$\sigma$, respectively. After removal of these two points, 158 measurements are left, spread over three years. A new CLEAN analysis revealed the strongest peak near 3.020 d, which is the same as the period from the ground-based observations. A least-squares fit gives $P_{\rm Hipparcos} = 3.01899 \pm 0.00064$ d with amplitude 3.5 $\pm$ 0.4 mmag, very similar to the values of \cite{jerzykiewicz:1992}.
A combined fit of ground- and space-based photometry, after normalising the average magnitude of the two datasets where we only used the Str\"{o}mgren $y$ (comparable to the $H_{\rm p}$ magnitude), gives as the best-fit photometric period $P_{\rm photometric} = 3.01938 \pm 0.00022$ d with amplitude 3.1 $\pm$ 0.5 mmag. Maximum light occurs at:

\begin{eqnarray}
\nonumber T({\rm LTPV + Hipp}) = &{\rm HJD}\ 2447620.48\pm0.56 &\\
&+\ n \times(3.01938 \pm 0.00022)&
\label{eq:phasephoto}
\end{eqnarray}

\begin{figure}[b!]
\epsfig{file=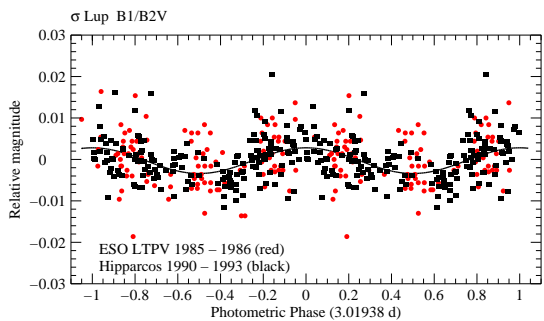,width=0.8\columnwidth,clip=}
\caption{LTPV with $y$ filter (red dots) and {\sl Hipparcos} (black squares) photometry as a function of phase with the best fit sinusoid overplotted. Estimated error bars are about 5 mmag.
}
\label{fig:photom}
\end{figure}

The best fit cosine function is overplotted on the phased data in Fig.\ \ref{fig:photom}.

As an attempt to reproduce the photometric light curve of \cite{jerzykiewicz:1992}, we conducted five nights of Str\"{o}mgren $v$ filter photometry of $\sigma$ Lup, using the 0.9~m CTIO telescope operated by the SMARTS\footnote{The Small and Moderate Aperture Research Telescope System (http://www.astro.yale.edu/smarts/)} consortium with a $2048 \times 2048$ CCD detector.  Although $\sigma$ Lup is a bright star, the associated errors, both instrumental and atmospheric, are at or beyond the amplitude of the reported variability.  We are thus unable with this instrumentation to contribute any further information regarding the photometric period.  Furthermore, $\sigma$ Lup lacks a suitable nearby comparison star, making differential photometry a difficult task.

\section{UV observations}
\label{observations}
An earlier UV observation with the OAO-2 satellite was reported by \cite{panek:1976}, who measured the equivalent widths of the \ion{Si}{iv} and \ion{C}{iv} doublets (4.5 \AA\ and 1.0 \AA).

\begin{figure}
\epsfig{file=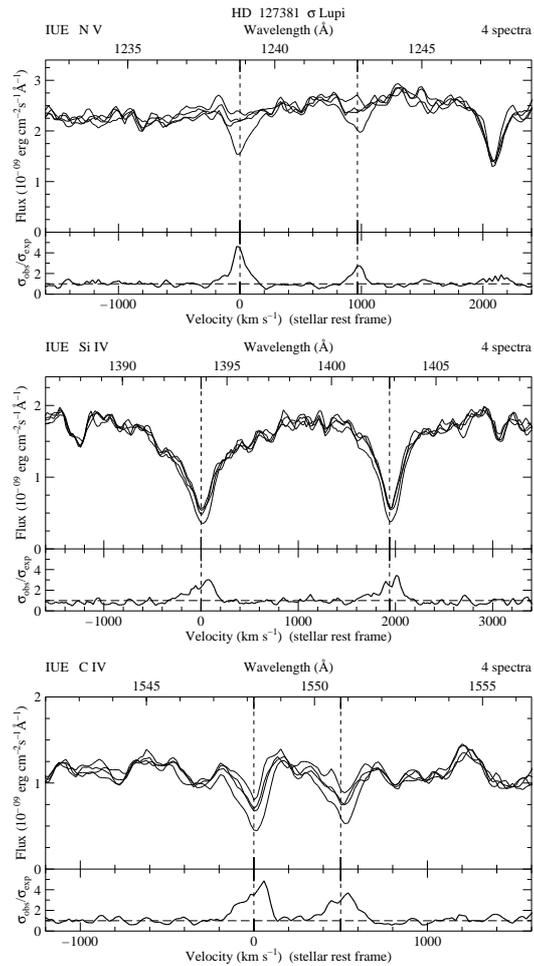,width=0.8\columnwidth,clip=}
\caption{Overplot of the four selected profiles of \ion{N}{v} ({\it top}), \ion{Si}{iv}
({\it middle}), and \ion{C}{iv} ({\it bottom}) of $\sigma$ Lup. The two doublet rest wavelengths are indicated by vertical dashed lines. Top scale: wavelength. Bottom scale: velocity with respect to the stellar rest frame.
In each panel the lower part displays the significance of the variability, expressed as the square root of the ratio of the measured to the expected variances. The changes are very similar to what is observed in known magnetic B stars (see Fig.\ \ref{fig:civ}), but not in other B stars. }
\label{fig:nsc}
\end{figure}

\begin{figure}[h!]
\epsfig{file=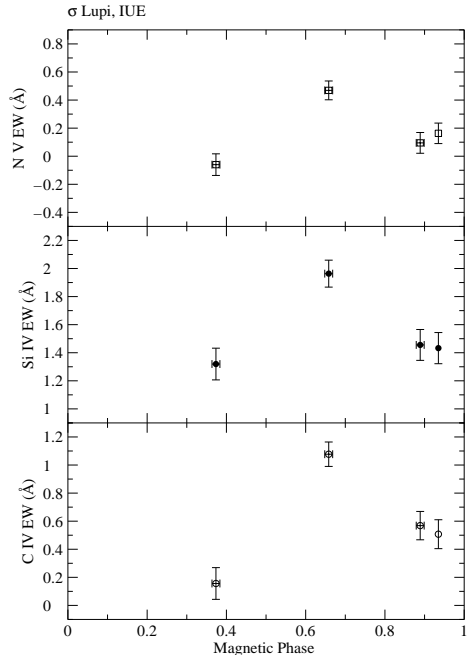,width=0.7\columnwidth,clip=}
\caption{Equivalent widths of the \ion{N}{v} ({\it top}), \ion{Si}{iv} ({\it middle}) and \ion{C}{iv}  ({\it bottom}) line profiles as a function of magnetic phase (Eq.~\ref{eq:magphase}).
}
\label{fig:ew}
\end{figure}

\begin{table*}
\caption[]{Journal of high resolution IUE observations of $\sigma$~Lup, and the measured equivalent widths.} 
\begin{tabular}{clcrccccc}
\hline
\hline
 Image & \multicolumn{1}{c}{Date} &HJD$^1$ & \multicolumn{1}{c}{$t_{\rm exp}$} &
Phase$^2$  & EW (\ion{N}{V}) (\AA)& EW (\ion{Si}{IV}) (\AA) & EW (\ion{C}{IV})  (\AA)\\
 SWP &  & $-$2440000 &    \multicolumn{1}{c}{s} & & [$-$250, 250] km\,s$^{-1}$  & [$-$300, 300] km\,s$^{-1}$& [$-$200, 200] km\,s$^{-1}$\\

\hline
 20261 & 1983-06-19 & 5505.242 & 109.6 & 0.637 & - & - & - \\
 22094 & 1984-01-25 & 5724.656 &  99.8 & 0.297 & - & - & - \\
 45218 & 1992-07-24 & 8827.829 &  45.6 & 0.935 &    0.163 $\pm$ 0.073 & 1.43 $\pm$ 0.11 & 0.51 $\pm$ 0.10 \\
 47374 & 1993-03-28 & 9074.610 &  49.8 & 0.658 &    0.469 $\pm$ 0.067 & 1.96 $\pm$ 0.10 & 1.08 $\pm$ 0.09 \\
 47881 & 1993-06-16 & 9155.283 &  49.8 & 0.374 & $-$0.060 $\pm$ 0.077 & 1.32 $\pm$ 0.11 & 0.16 $\pm$ 0.11 \\
 48225 & 1993-07-24 & 9193.076 &  47.8 & 0.889 &    0.095 $\pm$ 0.074 & 1.46 $\pm$ 0.11 & 0.57 $\pm$ 0.10 \\

\hline
\end{tabular}
\label{tab:results1}
\newline
1. Calculated at mid exposure. --
2. Rotational phase, calculated with Eq.~\ref{eq:magphase}.
\end{table*}

Table~\ref{tab:results1} contains a list of all available high-resolution UV spectra taken with the IUE
satellite between 1983 and 1993. The first two spectra had to be discarded in our analysis,
as they were severely overexposed, which prevented reliable results in the regions of the stellar
wind lines of interest because of the saturation and missing portions of the reduced spectra. 
Fig.~\ref{fig:nsc} shows overplots of regions around the resonance lines of \ion{N}{v}, \ion{Si}{iv}, and
\ion{C}{iv}, along with the significance of their variability. The \ion{C}{iii} complex near 1175 \AA\ also shows similar variability but in this wavelength region the signal-to-noise ratio (S/N) is too low to derive quantitative results. No other region in the
spectrum between 1150 and 1900 \AA\ showed significant variability. We sampled the spectra upon a uniform grid with 0.1 \AA\ spacing (degrading the resolution
somewhat), and locally transformed the wavelength into a velocity scale in the stellar rest frame, while correcting
for the (small) radial velocity of the star. The wavelengths of the strongest member of the doublet lines
were taken as the zero point. We normalised the average flux of the whole spectrum by using only portions
of the spectrum which were not affected by stellar wind variability or by echelle order overlap mismatches.
The overall flux level did not differ more than 2$\%$ for the 4 spectra, likely within the expected
photometric uncertainty of the instrument. We note that the absolute fluxes agree well with those measured with the OAO-2 satellite \citep{code:1979}. The lower panels in the figures display the quantity
$\sigma_{\rm obs}/\sigma_{\rm exp}$, which is the ratio of the observed to the expected value of sigma, the
square root of the variance. The expected value of $\sigma$ was calculated with a noise model as a
function of the flux $f$ applicable for high-resolution IUE spectra  \citep{henrichs:1994} in the form of
$A$\,tanh$(f/B)$ with best-fit parameters $A$ = 19, representing the maximum signal to noise ratio, and $B = 7.8 \times 10^{-10}$ erg cm$^{-2}$s$^{-1}$\AA$^{-1}$, representing the average of all flux levels outside the regions with variability (intrinsic or instrumental).

All doublet lines show significant variations in the velocity interval [$-$200, +200] km\,s$^{-1}$. For
all three ions the absorption line profiles in image number SWP 47374 are much deeper than in the other
spectra, whereas the lines in SWP 47881 have the highest relative flux. This parallel behaviour excludes an
instrumental effect as the origin. It exclusively occurs in magnetic B stars, like the examples in Fig.~\ref{fig:civ}, which made this star a prime candidate for magnetic measurements.

Table~\ref{tab:results1} gives the equivalent-width measurements of the three wind lines over the velocity interval indicated. The listed values are the sum of the equivalent widths of the two doublet members. The continuum was taken at the averaged flux level of the two end points of the interval. The error bars were calculated with the method of \cite{chalabaev:1983}.

The calculated phases of the IUE observations are given in column 5 of Table~\ref{tab:results1}, where we used Eq.~\ref{eq:magphase} below, in which phase 0 corresponds to maximum (positive) magnetic field, and apparently also maximum light (see Sect.\ \ref{sec:phaseanalysis}). From the figure it is obvious that all lines behave in the same manner, and that the two points near phase 0.9 have very similar values, which we interpret as a confirmation of the periodicity.  Depending on the inclination angle $i$ and the magnetic obliquity $\beta$,  a single or double sine wave with possibly unequal minima would be expected to fit the phased data, but with four points this is impossible to verify. The lowest absorption (or highest emission, as the lines become filled in) appears around phase 0.4, i.e. compatible with minimum light. Based on the UV behaviour of other hot, magnetic stars, this would occur when a magnetic pole points to the observer. This is opposite to what we would naively expect for a lightcurve with a single minimum. We expect that maximum absorption occurs when the magnetic equatorial plane, where most of the circumstellar material will be collected, intercepts the line-of-sight to the star.

\section{Magnetic observations and properties}

\label{magneticobs}
\begin{table*}
\caption[]{
Journal of observations and results of magnetic measurements of $\sigma$~Lup.
} 
\begin{tabular}{lclccrcrrrrr}
\hline
\hline

Nr. & Obs.$^1$& \multicolumn{1}{c}{Date$^2$} &HJD$^3$ & $t_{\rm exp}$ & \multicolumn{1}{c}{S/N$^4$} &
   Phase$^5$ & \multicolumn{1}{c}{$B_{l}\,^6$} & $\sigma(B_{l})$
   & \multicolumn{1}{c}{$N_{l}\,^7$} & $\sigma(N_{l})$\\
 &    &                 &$-$2450000   &  s        & \multicolumn{1}{c}{pxl$^{-1}$}  
   &    &\multicolumn{1}{c}{G} & G
   & \multicolumn{1}{c}{G} & \multicolumn{1}{c}{G}\\ 
\hline
1  & A & 2007-12-28 & 4463.24031 & 4$\times$400 &  880 & 0.141 &     79 &  20 &   4 & 20\\
2  & A & 2008-01-01 & 4467.25522 & 4$\times$400 &  620 & 0.470 &  $-$69 &  27 &  $-$4 & 27\\
3  & A & 2008-12-15 & 4816.25084 & 4$\times$400 &  720 & 0.043 &     75 &  25 &   8 & 25\\
4  & A & 2008-12-16 & 4817.24836 & 4$\times$400 &  720 & 0.373 &  $-$39 &  24 &  $-$7 & 24\\
5  & A & 2008-12-17 & 4818.24066 & 4$\times$400 &  360 & 0.702 &     95 &  49 &   0 & 49\\
6  & C & 2009-02-17 & 4880.12491 & 4$\times$130 & 1015 & 0.195 &     69 &  24 &  22 & 24\\
7  & C & 2009-02-17 & 4880.13320 & 4$\times$130 & 1188 & 0.198 &     96 &  21 &  $-$4 & 21\\
8  & C & 2009-02-17 & 4880.14136 & 4$\times$130 & 1140 & 0.201 &     52 &  22 &   9 & 22\\
9  & A & 2009-04-08 & 4930.25755 & 4$\times$400 &  750 & 0.797 &     57 &  24 & $-$23 & 24\\
10 & C & 2010-02-27 & 5255.10549 & 4$\times$200 & 1114 & 0.373 &  $-$70 &  22 &   1 & 22\\
11 & C & 2010-03-02 & 5258.09114 & 4$\times$200 & 1021 & 0.361 &  $-$15 &  24 & $-$37 & 24\\
12 & C & 2010-03-03 & 5259.08881 & 4$\times$200 &  779 & 0.692 &  $-$15 &  31 & $-$15 & 31\\
13 & C & 2010-03-04 & 5260.08395 & 4$\times$200 & 1099 & 0.021 &    117 &  22 & $-$20 & 22\\
14 & C & 2010-03-05 & 5261.10469 & 4$\times$200 &  959 & 0.359 &  $-$43 &  25 &  $-$6 & 25\\
15 & C & 2010-03-07 & 5263.05040 & 4$\times$200 & 1303 & 0.004 &    114 &  19 & $-$13 & 19\\
16 & C & 2010-03-07 & 5263.08299 & 4$\times$200 & 1326 & 0.014 &    103 &  19 &  $-$3 & 19\\
17 & C & 2010-03-08 & 5264.02698 & 4$\times$200 &  490 & 0.327 &  $-$28 &  50 &  26 & 50 \\
18 & C & 2010-03-08 & 5264.03882 & 4$\times$200 &  624 & 0.331 &  $-$30 &  39 &  27 & 39 \\
19 & C & 2010-03-08 & 5264.12734 & 4$\times$200 &  744 & 0.360 &  $-$31 &  33 & $-$40 & 33 \\
20 & C & 2010-05-28 & 5344.86081 & 4$\times$200 &  908 & 0.096 &     73 &  40 &   3 & 40 \\
21 & C & 2010-05-28 & 5344.87315 & 4$\times$200 &  579 & 0.100 &     14 & 186 &  28 &186 \\
22 & C & 2010-05-28 & 5344.88541 & 4$\times$200 &  664 & 0.104 &    173 &  62 &  75 & 62 \\
23 & C & 2010-05-30 & 5346.84258 & 4$\times$200 &  980 & 0.752 &     27 &  28 &  23 & 28 \\
24 & C & 2010-05-31 & 5347.83115 & 4$\times$200 & 1291 & 0.079 &    117 &  22 &  15 & 22 \\
25 & C & 2010-06-01 & 5348.88552 & 4$\times$200 &  870 & 0.428 & $-$123 &  29 &  52 & 29 \\
26 & C & 2010-06-02 & 5349.82460 & 4$\times$200 &  963 & 0.739 &      1 &  26 &  11 & 26 \\
\hline									  
\label{tab:magnetic}
\end{tabular}								  
\newline 1. A = obtained with the 
SEMPOL spectropolarimeter attached to the UCLES at the AAT; C = with ESPaDOnS attached to the CFHT. -- 2. Date when the first observation started. --
3. Calculated at the center of the total exposure time
($t_{\rm exp}$) of column 5. --
4. Quality of the Stokes $V$ spectra,
expressed as the signal to noise ratio per pixel around 550 nm in the reduced spectrum. --
5. Rotational phase, calculated with Eq.~\ref{eq:magphase}.-- 6. Longitudinal magnetic field values with their 1-$\sigma$ uncertainties. -- 7. Computed magnetic values of the diagnostic null (or $N$) spectrum with their 1-$\sigma$ uncertainties.
\end{table*}


Initial spectropolarimetric observations in both left- and right-hand circularly polarised
light were obtained at the Anglo-Australian Telescope (AAT) at six observing epochs,
as shown in Table \ref{tab:magnetic}. These were obtained using the SEMPOL
spectropolarimeter \citep{semel:1989, donati:2003} on the AAT in conjunction with the
University College London \'{E}chelle Spectrograph (UCLES). SEMPOL was positioned at
the AAT's f/8 Cassegrain focus and consists of an aberration-free beam splitter and
an achromatic $\lambda/4$ plate orientated to transmit circularly polarised
light. SEMPOL splits the observed light into left-hand and right-hand circular
polarisation and feeds it down twin fibres, which is fed into a Bowen-Walraven image
slicer at the entrance to the high-resolution UCLES spectrograph. The $\lambda/4$
plate can be rotated from $+45^{\circ}$ to $-45^{\circ}$ with respect to the beam splitter to
alternate the polarisation in each fibre. In practice, four subexposures are obtained, alternating
the wave plate orientation between $\pm 45\degr$. These subexposures are combined in
such a way as to eliminate spurious contributions to the polarisation spectrum. Further information on the
operation of the SEMPOL spectropolarimeter is provided by \citet{semel:1993}, \cite{semel:1989}
and \citet{donati:1997, donati:2003}.

The detector used was a deep depletion EEV2 CCD with $2048\times2048\times13.5\mu$m square
pixels. The UCLES was used with a 31.6 gr/mm grating covering 46 orders (\#84
to \#129). The central wavelength was set to $\sim$526.8 nm with a 
coverage from $\sim$437.7 nm to $\sim$681.5 nm with a resolution of $\sim$ 71000.

\begin{figure}[h!]
\epsfig{file=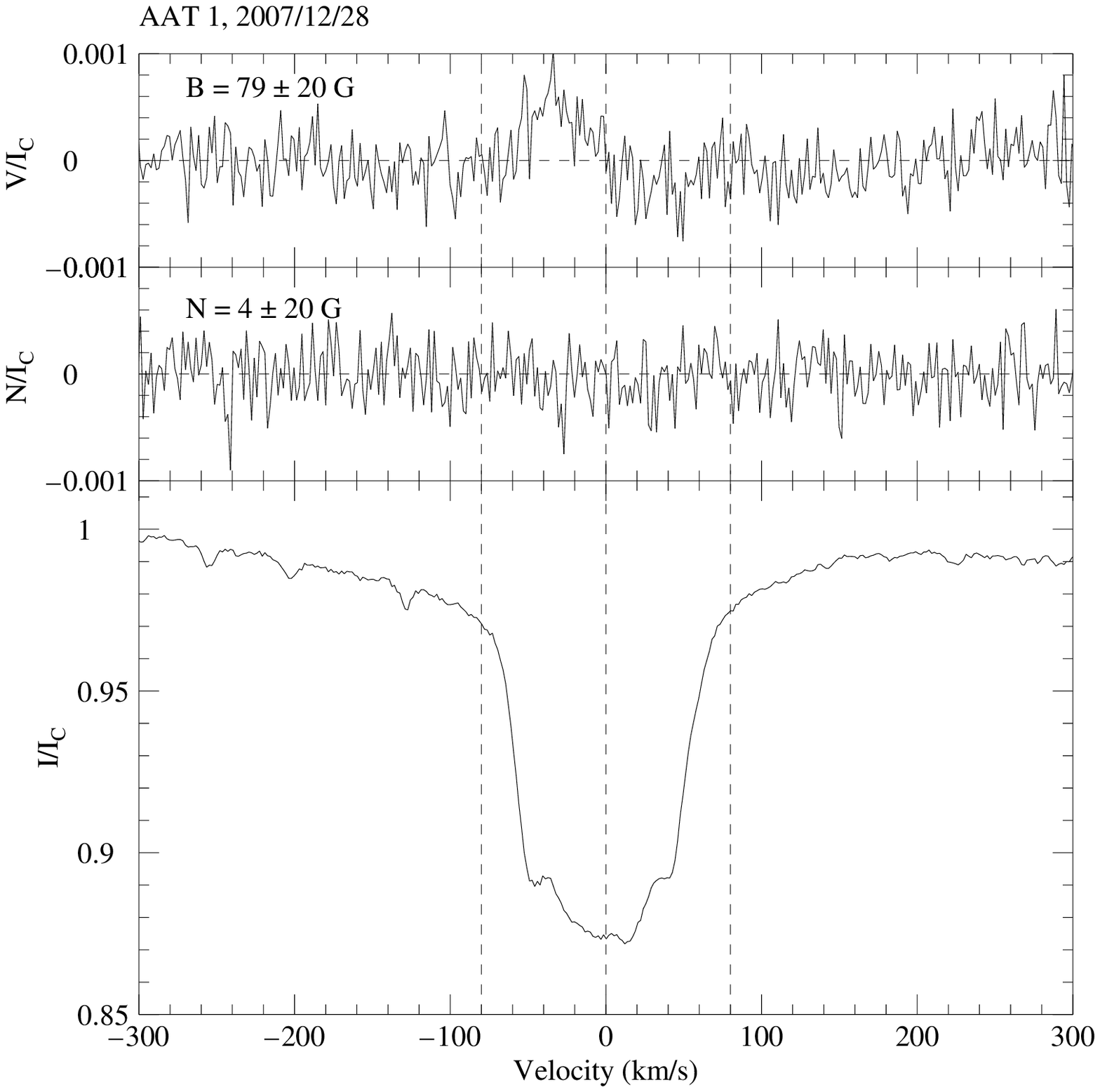,width=0.75\columnwidth,clip=}
\epsfig{file=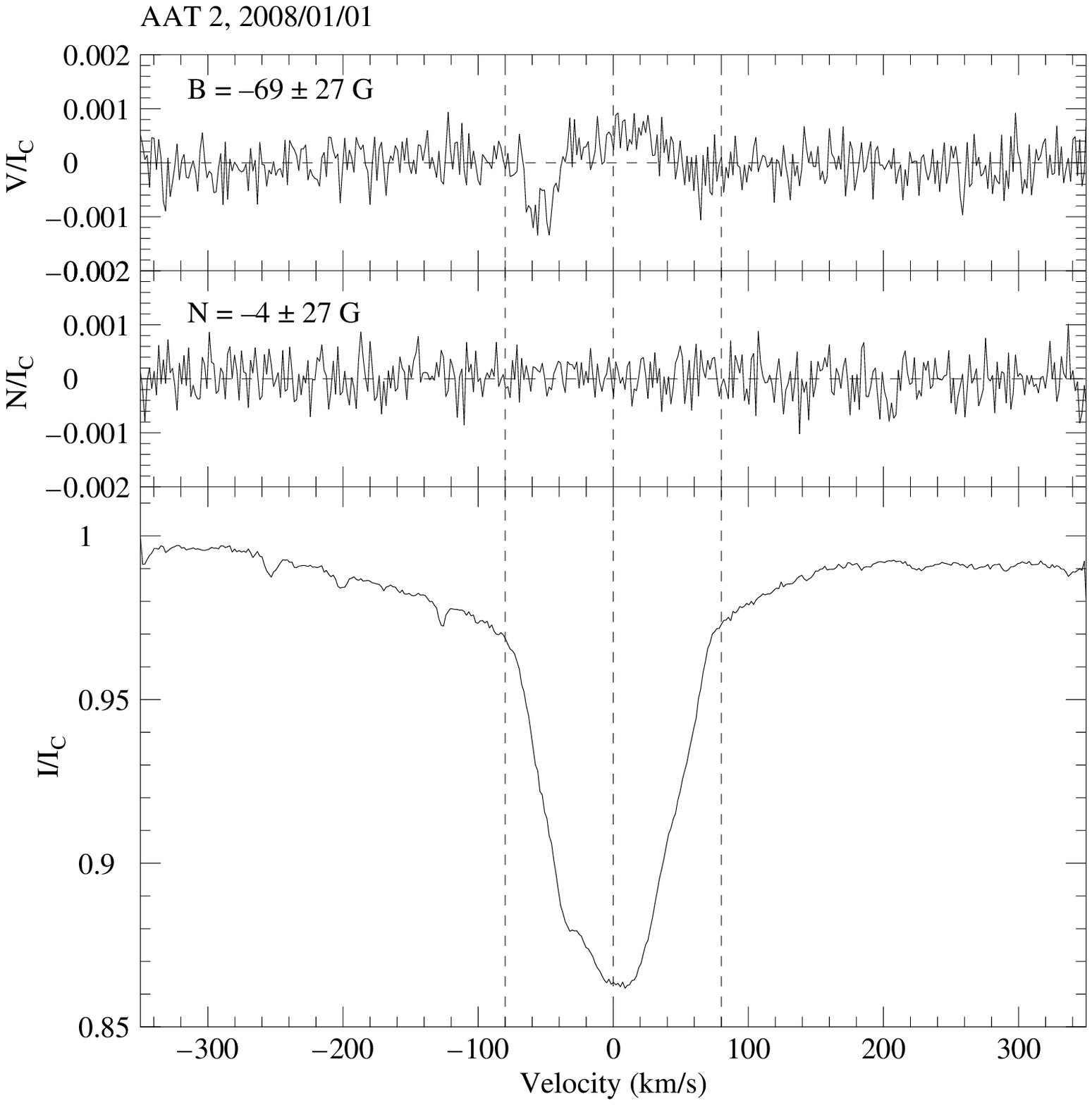,width=0.75\columnwidth,clip=}
\caption{Discovery observations of the Zeeman signature in $\sigma$ Lup at the AAT on December 28, 2007, (top) and four days later (bottom), at magnetic phases 0.14 and 0.47, respectively. Each figure shows the LSD Stokes unpolarised $I$ (lower panel) and circularly polarised $V$ (top panel) profiles of SEMPOL spectra, all normalised by the continuum.  The middle panel displays the (null) $N$ profile for integrity purposes. The integrated Zeeman signature in the $V$ profile over the width of the line (between the outer vertical dashed lines) gives the value of the longitudinal component of the magnetic field, integrated over the stellar surface. On the first night we obtained $B_l = 79 \pm 20$ G, whereas on the second night the $V$ signature changed sign, and we obtained $B_l = -69 \pm 27$ G. Note the irregular bumps in the $I$ profiles, which indicate the presence of spots on the stellar surface.}
\label{fig:lsd}
\end{figure}


Prompted by the magnetic detection in these spectra (see Fig.~\ref{fig:lsd}), subsequent measurements were obtained with ESPaDOnS at the Canadian-France-Hawaii Telescope (CFHT), as part of the Magnetism in Massive Stars (MiMeS) Large Program. The ESPaDOnS spectropolarimeter and observing procedure are fundamentally similar to the SEMPOL/UCLES combination, providing circularly polarised spectra with a resolving power $R\sim 65000$ from $\sim$$370$ to $1000$ nm. 
A total of 6 AAT spectra and 20 CFHT spectra were acquired, with peak S/N pxl$^{-1}$ ranging from 360--880 (AAT) and $\sim$$500$--$1300$ (CFHT). For the log of observations see Table \ref{tab:magnetic}.

Optimal extraction of the SEMPOL and ESPaDOnS \'{e}chelle spectropolarimetric observations was done using
the {\small ESpRIT} and {\small Libre-ESpRIT} (\'{E}chelle Spectra Reduction: an Interactive Tool;
\citealt{donati:1997}) reduction packages. Preliminary processing involved removing the bias and flat-fielding using a
nightly master flat. Each spectrum was extracted and wavelength calibrated against a Th-Ar lamp, then normalised to the continuum using polynomial fits to individual spectral orders.

\subsection{Spectropolarimetric analysis using LSD}

As the Zeeman signatures in atomic lines are typically extremely small (typical relative
amplitudes of 0.1 per cent or less), we have applied the technique of Least-Squares
Deconvolution (LSD, \citealt{donati:1997}) to the photospheric spectral lines in
each \'{e}chelle spectrum in order to create a single high-S/N
profile for each observation. In order to correct for the minor instrumental shifts
in wavelength space, each spectrum was shifted to match the Stokes {\it I} LSD profile of the telluric lines contained in the spectra, as described
by \citet{donati:2003} and \citet{marsden:2006a}. The line mask used for the stellar LSD was
a B2 star linelist containing 170 He and metal lines created from the Kurucz atomic database and ATLAS9 atmospheric
models \citep{kurucz:1993} but with modified depths such that they are proportional to 
the depth of the observed lines.
Further information on the LSD procedure is provided by \citet{donati:1997} and \citet{kochukhov:2010}.

\section{Spectropolarimetry: results}
\label{magneticresults}

The presence of a magnetic field on a star produces a signal in the Stokes $V$ profile.
The Stokes $V$ profile is created by constructively combining the left- and right-hand
circularly polarised light together from the 4 sub-exposures, while a diagnostic null ($N$) profile
was created by destructively combining the 4 sub-exposures. This null profile was used
to check if any false detections have been introduced by variations in
the observing conditions, instrument or star between the sub-exposures. Fig.\ \ref{fig:lsd} shows the
Stokes $V$ LSD profiles for the two discovery observations, which show opposite Zeeman signatures. 
As mentioned earlier, the Stokes $V$
signature is much smaller than Stokes $I$, typically less than 0.1\% of the
continuum. By combining the more than 170 lines, the S/N is improved
dramatically. We used the same line list for all spectra. Table \ref{tab:magnetic} includes the resulting mean
SNR for each of the observations.  The observing conditions
were often adverse, particularly during the December 2008 run at the AAT as the star was very
low in the East and was taken just prior to $12{^\circ}$ twilight. Hence the detections on
the 15th and again on the 17th of December, 2008 were only marginal in LSD profiles sampled at the nominal resolution
of the instrument.  The observations at CFHT were all taken near airmass 3 because of the low declination ($-50^{\circ}$) of the star. The observation at JD 2455344.87415 (number 21 in Table \ref{tab:magnetic}) was taken under very poor weather conditions, leading to a veer large uncertainties (null spectrum with $N = 28 \pm 186$ G, and $B = 14 \pm 186$ G). This observation was disregarded in further analysis.

\subsection{Longitudinal magnetic field}

\begin{figure}[t!]
\epsfig{file=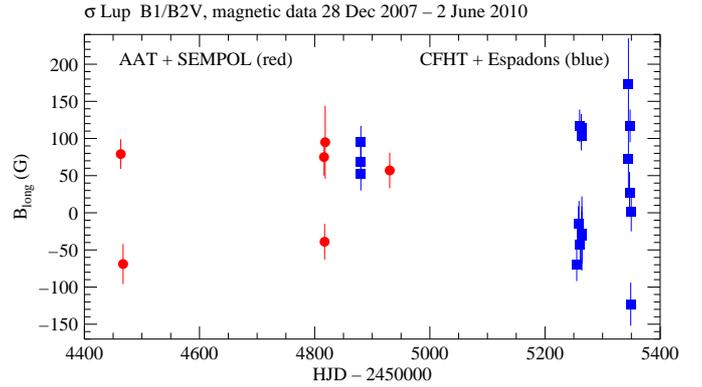,width=\columnwidth,clip=}
\caption{Magnetic data as a function of time. Observations from AAT (red dots) and CFHT (blue squares) are indicated.   
}
\label{fig:magneticobs}
\end{figure}

We derived the longitudinal component of the magnetic field, integrated over the visible 
hemisphere, from the Stokes $I$ and $V$ LSD profiles in the usual manner by calculating \citep{mathys:1989, donati:1997}:

\begin{equation}
\label{equ:bl}
B_{\ell} = (-2.14\times 10^{11}\,{\rm G})\frac{\int vV(v) dv}
{\lambda g c\int [1 - I(v)] dv},
\end{equation}

\noindent where $\lambda$, in nm, is the mean, S/N-weighted wavelength, $c$ is the velocity of
light in the same units as the velocity $v$, and $g$ is the mean, S/N-weighted value of the Land\'e
factors of all lines used to construct the LSD profile. We used $\lambda$ =
515 nm and $g$ = 1.27.
The noise in the LSD spectra was measured and is given in
Table~\ref{tab:magnetic}, along with the S/N obtained in
the raw data.  The integration limits were chosen at $\pm$ 90 km\,s$^{\rm -1}$. 
The longitudinal field is measured with a typical 1$\sigma$ uncertainty of 20--30~G, and varies between about $+100$ and $-100$ G. In contrast, the same quantity measured from the $N$ profiles show no significant magnetic field. The longitudinal field data as a function of time are shown in Fig.~\ref{fig:magneticobs}, and are reported in Table 3.

\subsection{Magnetic period and phase analysis}
\label{sec:phaseanalysis}
\begin{figure}[h!]
\epsfig{file=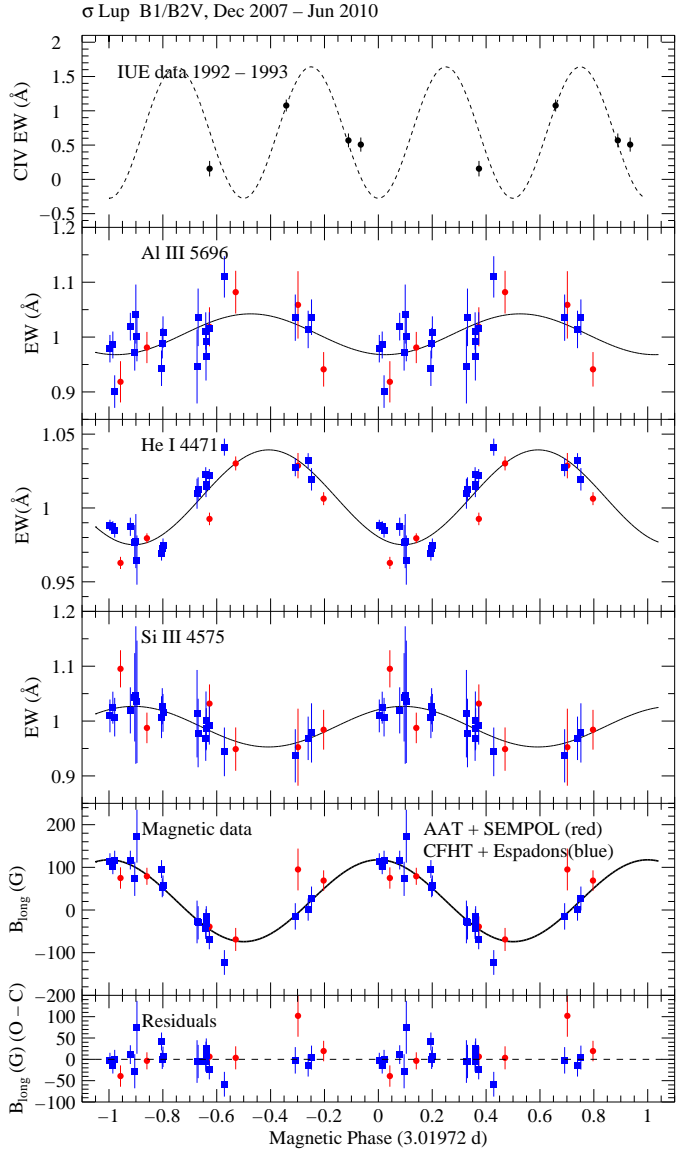,width=\columnwidth,clip=}
\caption{Magnetic and equivalent-width data (red dots: AAT; blue squares: CFHT) phased with Eq.~\ref{eq:magphase}, repeated over two rotational periods. Lower two panels: magnetic data with their residuals to the best-fit cosine curve. Panels 3, 4 and 5 from bottom: equivalent widths of the \ion{Si}{iii} 4575, \ion{He}{i} 4471 and \ion{Al}{iii} 5696 lines with a best-fit sinusoid superposed. Note the opposite behavior of the different lines. Top: \ion{C}{iv} 1540 equivalent width fifteen years earlier. The dashed curve is a suggested double sine wave, with arbitrary amplitude, and phased according to what is observed in other similar magnetic stars. 
}
\label{fig:magnetic}
\end{figure}

A best fit to the magnetic data in Table~\ref{tab:magnetic} of the form
\begin{equation}
B_\ell (t) = B_0+ B_{\rm max} \cos (2\pi(t/P+\phi))
\end{equation}
 with $1/\sigma(B)^2$ error bars as weights 
yields $P = 3.01972 \pm 0.00043$ d, $B_0 = 22 \pm 6$ G, $B_{\rm max}  = 96 \pm 8 $ G, and $\phi = 0.11 \pm 0.24$ with a reduced $\chi^2 = 1.01$. An overplot with the data, folded in phase, along with the residuals is presented in the two lower panels of Fig.~\ref{fig:magnetic}.  The maximum magnetic field value occurs at:
\begin{eqnarray}
\nonumber T({\rm B_{max}}) = &{\rm HJD}\ 2455106.01\pm0.72 &\\
&+\ n \times(3.01972 \pm 0.00043).&
\label{eq:magphase}
\end{eqnarray}

 The obtained period is very close to the photometric period of $3.01938 \pm 0.00022$ d (see Eq.~\ref{eq:phasephoto}). Extrapolating backwards to the epoch of photometric observations (with 2479 cycles), we find that the epoch of maximum light at HJD $2447620.48 \pm 0.56$ coincides within the errors with the epoch of expected maximum magnetic field at HJD $2447620.86 \pm 0.72$. As we have no a priori knowledge of why maximum light would occur at maximum field strength, we cannot derive a more accurate value of the rotational period from combining the phase information.

\subsection{Magnetic geometry}
The adopted values of the rotation period, $v$sin$i$ and stellar radius (see Table~\ref{tab:parameters}) imply that the inclination, $i$, of the rotation axis with the line of sight  is about $58^{\circ}$, but not well constrained between $43^{\circ}$ and $90^{\circ}$. For a centered dipole model, the magnetic tilt angle with respect to the rotation axis, $\beta$, is then constrained by the observed ratio $B_{\rm max}/B_{\rm min} = \cos(i+\beta)/\cos(i-\beta) = -1.59^{+0.58}_{-0.40}$, implying $\beta$ close to $90^{\circ}$. In other words, the line of sight is not so far ($\lesssim 45^{\circ}$) from the rotational equatorial plane, and per rotation we see both magnetic poles passing. Such a dipole geometry implies a polar field of $B_{\rm p} \simeq 500$ G.

\section{Equivalent widths}
\label{ew}
\begin{figure}[h!]
\epsfig{file=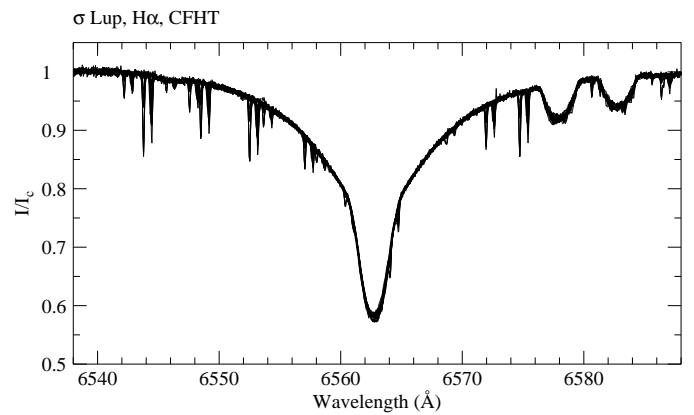,width=\columnwidth,clip=}
\caption{Overplot of spectra around H$\alpha$, showing no obvious emission, and no significant variability.}
\label{fig:ha}
\end{figure}

The equivalent widths were measured for the following 15 spectral lines: 
\ion{C}{ii} $\lambda$4267, \ion{He}{i}  $\lambda$4471, 4920, 6678,
\ion{N}{ii} $\lambda$4601 (blended with \ion{O}{ii}), 4630,   \ion{O}{ii}$\lambda$ 4416, 4591, \ion{Si}{ii} $\lambda$5040, 5055, \ion{Si}{iii} $\lambda$4552 (blended with \ion{N}{ii}), 4568, 4575, \ion{Al}{iii} 5696, and
H$\alpha$ $\lambda$6563.  The variation as a function of rotational phase of three representative lines are shown in Fig.~\ref{fig:magnetic}. 
The three helium lines show the most significant modulation, in antiphase with the magnetic field, whereas opposite phase behavior, i.e.\ in phase with the magnetic field, is found among the
 five silicon lines. The slightly variable \ion{Al}{iii} varies in phase with the helium lines. The nitrogen lines were not detected to vary.  None of the other studied lines show significant periodicity. This includes H$\alpha$, which did not show any significant emission, see Fig.~\ref{fig:ha}.  We note that the antiphase behaviour between the He and Si lines is similar to what is observed in other magnetic He-strong and He-weak  B stars, like for instance HD 37776, for which \cite{khokhlova:2000} showed that helium is overabundant in regions where silicon is underabundant, and vice versa.  The surface abundances of some of these elements are apparently not homogeneously distributed over the stellar surface, causing a rotational modulation of the equivalent widths, and in total luminosity. The irregular bumps in the $I$ profiles in Fig. \ref{fig:lsd} are therefore likely explained by the presence of spots for a number of elements, associated with the magnetic field configuration. Zeeman Doppler Imaging techniques may give a more quantitative result.

\section{Abundance determination}
\label{abund}

For its spectral type it is not unusual that this star is magnetic: in fact it could be a He peculiar star.  The ultraviolet \ion{He}{ii} line at 1640.41 \AA\ is not prominent. 
Many magnetic early B stars have abundances deviating from solar. 

 The available spectra include lines of H, He, C, N, O,
Ne, Mg, Si, S, and Fe. The spectral lines of most of
the elements are mildly variable, particularly in the shape and
depth of the cores, but very little in the line wings.
This strongly suggests that the atmospheric chemistry of $\sigma$ Lup
is not uniform, but is somewhat patchy. The fact that the overall
equivalent width of most lines remains approximately constant in spite
of the small core variations suggests that this patchiness is rather
small scale -- there is no strong evidence for variations
from one magnetic hemisphere to the other, except for a mild variation
of helium.

To obtain a first characterisation of the atmospheric chemistry of the
star, mean abundances have been determined for several elements from
each of two spectra. Such abundances, for elements that are certainly
non-uniform on a small scale, correspond roughly to hemispheric
average abundances for the hemisphere producing the observed spectrum.
Two spectra have been analysed, one from near the negative field
extremum and He line strength maximum at phase 0.428 (obtained on
2010-06-02), and one from near field maximum and He minimum at phase
0.195 (obtained on 2009-02-17).

The abundance determinations have been made using the line synthesis
program ZEEMAN \citep{landstreet:1988,wade:2001}, designed to compute line profiles for magnetic stars from a simple
model of magnetic field structure (a low order multipole expansion)
and a simple distribution model which in this case is taken to be a
uniform distribution over the star. It takes as input a 
model atmosphere in local thermodynamic equilibrium  (LTE)
computed with Kurucz' ATLAS for $T_{\rm e} = 23000$~K and $\log g =
4.0$, and computes the observed spectrum. The program determines a
best value of radial velocity $v_{\rm r} = 0.0 \pm 0.5$~km\,s$^{-1}$,
rotational velocity $v \sin i = 68 \pm 6$~km\,s$^{-1}$, and abundance
of a single element at a time by optimising the fit to a segment of
observed spectrum (see e.g.\ \citealt{bagnulo:2004}), iterating this
process until a satisfactory fit. A dipolar field with polar strength $B_{\rm d} = 400$ G observed
at about $45^\circ$ has been assumed, but the results are essentially
independent of the assumed field geometry.

\begin{figure}[b!]
\epsfig{file=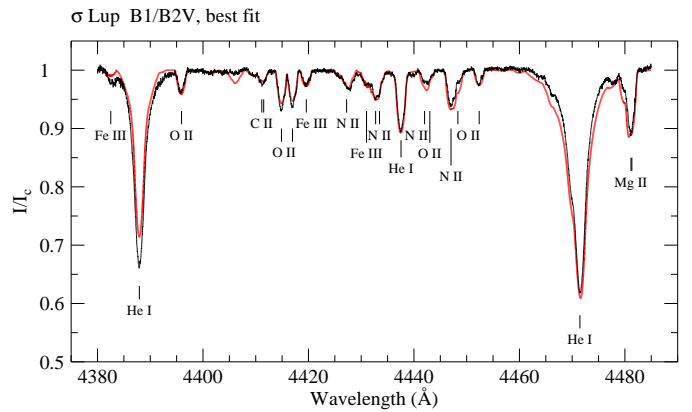,width=\columnwidth,clip=}
\caption{Portion of the spectrum  at phase 0.428, with most lines identified, overplotted with the model fit to determine abundances.
}
\label{fig:abund}
\end{figure}

The program assumes LTE. Although this is not a good approximation in
general for a star of $T_{\rm e} = 23000$~K,  \cite{przybilla:2011} showed that LTE modelling can yield
meaningful abundances up to about this effective temperature, provided
the right spectroscopic indicators are employed. They provide a list
of some of the indicators to use. We have identified others by
synthesising an available spectrum of \object{$\alpha$~Pyx} = \object{HD 74575}, a star
very similar to $\sigma$~Lup, for which the fundamental
parameters and abundance table are known with high accuracy \citep{przybilla:2008}.
By creating a synthetic spectrum of $\alpha$~Pyx based
on these stellar parameters, we can identify spectral lines whose
strengths are correctly computed in LTE, and then use these lines to
determine abundances in $\sigma$~Lup.

Abundances were determined using several $\sim$$100$~\AA\ windows. A
sample of the fit for phase 0.428 is
shown in Fig.~\ref{fig:abund}. The results are summarised in Table~\ref{abundances}, which
lists for each element studied both the logarithmic value of the ratio
of the number density of the element to the number density of H for
each of the two spectra, the value of this ratio in the Sun as
reported by \cite{asplund:2009}, and the logarithmic ratios of the
values in $\sigma$~Lup relative to the Sun (average values for
elements that have closely similar abundances in the two spectra).

\begin{table}
\caption{Characteristic chemical abundances of $\sigma$ Lup, determined at rotational phases near magnetic field extrema (columns 2 and 3),
and compared with solar values (column 4).
}
\begin{tabular}{c@\ c c c@\ c@\ }
\hline\hline
Element & $\log\left(\frac{n_{\rm X}}{n_{\rm H}}\right)_{\rm 0.195}$ & $\log\left(\frac{n_{\rm X}}{n_{\rm H}}\right)_{\rm 0.428}$ & $\log\left(\frac{n_{\rm X}}{n_{\rm H}}\right)_{\odot}$ &  $[X/H]$ \\
\hline
He         &  $-0.88  \pm 0.12$    &  $-0.75  \pm 0.14$    &   $-1.07$     &   $+0.19$ to $+0.32$ \\
C          &  $-3.90  \pm 0.19$    &  $-3.85  \pm 0.15$    &   $-3.57$     &   $-0.30$ \\
N          &  $-3.73  \pm 0.16$    &  $-3.75  \pm 0.15$    &   $-4.17$     &   $+0.43$ \\
O          &  $-3.28  \pm 0.16$    &  $-3.33  \pm 0.14$    &   $-3.31$     &   $+0.00$ \\
Mg         &  $-4.57  \pm 0.15$    &  $-4.52  \pm 0.15$    &   $-4.40$     &   $-0.15$ \\
Al         &  $-5.80  \pm 0.15$    &  $-5.75  \pm 0.15$    &   $-5.55$     &   $-0.22$ \\
Si         &  $-5.0   \pm 0.4$     &  $-5.0   \pm 0.4$     &   $-4.49$     &   $-0.5 $ \\
S          &  $-5.68  \pm 0.15$    &  $-5.68  \pm 0.15$    &   $-4.88$     &   $-0.80$ \\
Fe         &  $-4.60  \pm 0.10$    &  $-4.60  \pm 0.10$    &   $-4.50$     &   $-0.10$ \\
\hline
\label{abundances}
\end{tabular}
\end{table}

This table needs several remarks. Uncertainties in
abundance are estimated from the scatter in the value derived from
different spectral windows. This scatter probably arises in part from
inaccuracy in atomic data, in part from probably not precisely correct
fundamental parameters, in part from neglect of non-LTE effects, and
in part from the small-scale abundance inhomogeneities that are not
modelled.

For He, we have used the relatively weak \ion{He}{i} lines at  4437, 4713, 5015, and 
5047~\AA, which do not have signficant forbidden 
components, and for which LTE and non-LTE
calculations give very similar profiles at 22000~K \citep{przybilla:2011}.
We also modelled the very weak \ion{He}{i} line
at 4686~\AA; this line is approximately the correct strength when
computed with the abundance of He derived from the \ion{He}{i} lines,
consistent with the values of $T_{\rm e}$ and $\log g$ in Table \ref{tab:parameters}. 
The He abundance in both spectra is weakly overabundant with respect to the
solar value, by about a factor of two, so $\sigma$~Lup is a mild He-rich
star. An abundance difference of about 0.1--0.15 dex is consistently
found between the two spectra for all the lines modelled, with a
larger abundance in the spectrum at phase 0.428, consistent with the
larger equivalent width observed for \ion{He}{i} 4471 (Fig.~\ref{fig:abund}). 

\cite{przybilla:2011} point out that many weak lines of \ion{N}{ii}
and \ion{O}{ii} are well described by LTE, but that \ion{C}{ii} lines
are mostly out of LTE. For C they recommend using the lines of \ion{C}{ii}
 multiplet (16) around 5145~\AA\ for LTE abundance determination.
Accordingly, the abundances of N and O are based on several weak lines
of each element in several spectral window between 4400 and 5050~\AA,
while those of C are based on the recommended multiplet. The
uncertainty is estimated from the dispersion of the abundances found
in various spectral windows. From these data we find that there is no
significant difference in the abundance of these three elements
between the two magnetic hemispheres. C is found to be underabundant
with respect to the Sun by about a factor of 2; N is overabundant by
about a factor of 2, and O is approximately solar.
Abundances of Mg, Al, S, and Fe are determined using weak lines of the
dominant ionisation states. None seems to be significantly variable,
and are all mildly subsolar in abundance, except for S which is almost
one dex underabundant.

A number of lines of \ion{Si}{ii} and \ion{}{iii}, and a few lines of
\ion{Si}{iv}, are visible in the spectra of $\alpha$~Pyx and
$\sigma$~Lup. When these Si lines are synthesised for $\alpha$~Pyx in
LTE using the parameters and abundances of this star, most of the
observed lines are found to be very discordant with the computed
spectrum. The same problem is present in the spectrum of
$\sigma$~Lup.  The abundance of Si in $\sigma$~Lup was initially
obtained using the weak \ion{Si}{ii} lines at 5040 and 5056~\AA, and
the considerably stronger \ion{Si}{iii} triplet 4552, 4567, and
4574~\AA. The two levels of ionisation gave abundances that differed
by about a factor of 30, from about $-5.50$ for the \ion{Si}{ii} lines
to $-3.95$ for the \ion{Si}{iii} lines. From the discussion of this
element by \cite{przybilla:2011}, it appears that this is largely a
problem of major departures of Si levels from LTE, and that even
non-LTE calculations (e.g. \citealt{becker:1990a}) do not fully
account for the departures.

Accordingly, we have simply searched the spectrum of $\alpha$~Pyx for
lines of Si that are approximately correct with the Si abundance found
by \cite{przybilla:2008}. The most useful lines seem to be \ion{Si}{iv} at 4088 and 4116~\AA. These lines are consistent with
an abundance of Si that is about a factor of 4 below the solar value,
in both the spectra modelled. However, we give this element an
unusually large uncertainty in view of the exceptional difficulties of
LTE analysis. Although the exact value is extremely uncertain, it
appears that the abundance of Si in $\sigma$~Lup is probably no
higher than the solar ratio.  As expected, line profiles of most
\ion{Si}{ii} and \ion{}{iii} lines computed with this abundance do not
match the observed profiles at all well.

\section{Discussion and conclusions}

In this paper we predicted and confirmed that $\sigma$ Lup is an oblique magnetic rotator. 
The period determined from photometric measurements is $3.01938\pm 0.00022$~d, while that determined from the magnetic measurements is $3.01972\pm 0.00043$~d. These periods are in formal agreement, and we interpret them to be the rotational period of the star. Using the $\pm 100$~G variation of the longitudinal magnetic field, in combination with our estimate of the rotation axis inclination $i\simeq 58\degr$, we model the magnetic field as a dipole with obliquity $\beta\simeq 90\degr$ and surface polar field strength $B_{\rm d}\simeq 500$~G. Modeling of the spectrum using the LTE synthesis code ZEEMAN reveals enhanced chemical abundances of a variety of elements. We observe variations of the equivalent widths and shapes of line profiles that repeat with the rotation period, indicating that chemical elements are distributed non-uniformly across the stellar surface, with patterns that depend on the element in question.

These properties do not stand out in any particular way from the general population of early B-type magnetic stars. The longitudinal field variation is approximately sinusoidal, indicative of an important dipolar component to the field. The polar field strength is not atypical for such stars, nor is the large inferred value of the obliquity \citep[see, e.g.][]{2011MNRAS.412L..45P, 2011A&A...536L...6A}. The rotational period is within the range of periods observed for such stars, and the chemical peculiarities and their non-uniform distributions are similar to those of other stars with similar temperatures. We therefore conclude that $\sigma$~Lup is a rather normal magnetic B-type star.

As summarised by \cite{jerzykiewicz:1992}, similar photometric periods have been found in a number of Be and Bn stars \citep{balona:1987, vanvuuren:1988, cuypers:1989, balona:1991b}, see also \cite{carrier:2003}, termed \object{$\lambda$ Eri} variables. These are short-term periodic Be stars with strictly periodic variations with periods in the range 0.5--2.0 d. \cite{Vogt:1983} suggested that these variations could be explained either by non-radial pulsations or by clouds close to the stellar photosphere.
\cite{balona:1991a} and \cite{balona:1991b} suggested a rotational origin, specifically caused by material trapped above the photosphere by magnetic fields. In this aspect it is interesting to mention that \cite{neiner:2003b} reported a possible detection of a magnetic field in the Be star \object{$\omega$ Ori}, which also belongs to this class (although recent results are unable to confirm the reported field: see \cite{neiner:2012} .  As was demonstrated by e.g. \citet{2009A&A...499..567K} and assumed in this paper, spots caused by surface abundance inhomogeneities, associated with the magnetic field, cause the periodic photometric variations.  Further investigation is needed to establish whether a similar phenomenon holds for $\lambda$ Eri stars as well.

This star is the fourth confirmed B star which has been diagnosed to be magnetic from the variability of its ultraviolet stellar wind lines, i.e.\ together with $\beta$ Cep, V2052 Oph and $\zeta$ Cas. These stars, however, had their rotation period established from UV data, whereas the period of $\sigma$ Lup was found from photometry. It could be due to the short rotation period that the photometric changes are more easily detectable in this star, and/or that surface spots are less pronounced in the other stars. 
From these four B stars the presence of a magnetic field apparently does not depend on the pulsation or rotation properties.

It is interesting to compare the abundances of these magnetic B stars, as reported by \cite{morel:2006}. All these stars are enhanced in nitrogen, whereas all but $\beta$ Cep are helium enriched. Apparently nitrogen enrichment in the presence of a magnetic field is even more common than helium over- or under-abundance. This property was in fact used to search for a magnetic field in $\zeta$ Cas. In this context we also note that in the previously-known nitrogen enriched B0.5V star $\xi^1$ CMa a magnetic field was found by \cite{hubrig:2006}, which was confirmed by \cite{silvester:2009}.   On the other hand, other some nitrogen enriched early B stars (e.g. $\delta$~Cet) are not found to be magnetic \citep{2011mast.conf...23W}. Thus N enrichment, while not necessarily a direct tracer of magnetism in B stars, could still serve as an additional indicator that will be useful to discover more magnetic B stars.

{\acknowledgements  We are grateful to Coralie Neiner and Nolan Walborn for their comments on the manuscript. HFH fondly acknowledge the warm hospitality of the Department of Astrophysics, Radboud University, Nijmegen, under the inspiring directorship of Prof. Paul Groot, where this work was initiated. The authors would like to thank the technical staff at the AAT for
their helpful assistance during the acquisition of the spectroscopic
data. GAW and JL acknowledge Discovery Grant support from the Natural Sciences and Engineering Research Council of Canada. This project used the facilities of SIMBAD and Hipparcos. This research made use of INES data from the IUE satellite and of NASA's Astrophysics Data System.}

\bibliographystyle{aa}
\bibliography{references}
\end{document}